\documentclass[reprint,superscriptaddress,
frontmatterverbose,amsmath,amssymb,aps,prc,floatfix]{revtex4-1}
\usepackage{graphicx} 
\usepackage{times}
\usepackage{braket}

\def\fun#1#2{\lower3.6pt\vbox{\baselineskip0pt\lineskip.9pt
 \ialign{$\mathsurround=0pt#1\hfil##\hfil$\crcr#2\crcr\sim\crcr}}}
\usepackage{dcolumn}
\usepackage{bm}

\usepackage{ulem}
\usepackage{color}
\definecolor{verde}{rgb}{0.0,0.7,0.0}

\begin{document}


\title{$\beta$-Delayed Neutron and Fission Calculations with Relativistic QRPA and Statistical Model}


\author{Futoshi Minato}
\email{minato.futoshi@jaea.go.jp}
\affiliation{Nuclear Data Center, Japan Atomic Energy Agency, Tokai, Ibaraki 319-1195, Japan}

\author{Tomislav Marketin}
\affiliation{Department of Physics, Faculty of Science, University of Zagreb, 10000 Zagreb, Croatia}

\author{Nils Paar}
\affiliation{Department of Physics, Faculty of Science, University of Zagreb, 10000 Zagreb, Croatia}

\date{\today}

\begin{abstract}
 \noindent
 \textbf{Background:} 
A role of $\beta$-delayed neutron emission and fission in $r$-process nucleosynthesis attracts a high interest. Although the number of study on them covering $r$-process nuclei is increasing recently, uncertainties of $\beta$-delayed neutron and fission are still large for $r$-process simulations.
\\
 \textbf{Purpose:} 
Our purpose is to make a new database on $\beta$-delayed neutron emission and fission rates. 
To this end, the data that are not investigated experimentally have to be predicted.
Microscopic theoretical approaches based on a nuclear energy density functional and statistical models are one of the competent tools for the prediction.
We developed a new theoretical framework for $\beta$-delayed neutron emission and fission, and apply it to make a new data table.
\\
 \textbf{Methods:} 
To obtain $\beta$ strength function, proton-neutron relativistic quasiparticle random phase approximation is adopted. 
Particle evaporations and fission from nuclear highly excited states are estimated by the Hauser-Feshbach statistical model. 
$\beta$-delayed neutron branching ratios $(P_{n})$ are calculated and compared with experimental data.
$\beta$-delayed fission branching ratio ($P_f$) are also assessed by using four different fission barrier data.
\\
 \textbf{Results:} 
Calculated $P_{n}$ values are in a good agreement with experimental data and the root mean square deviation is comparable to results of preceding works.
It is found that energy withdrawal by $\beta$-delayed neutron emission sensitively varies $P_n$ values for nuclei near the neutron drip line.
$P_{f}$ are sensitively dependent on fission barrier data.
\\
\textbf{Conclusions:} 
Newly calculated data on $\beta$-delayed neutron emission and fission are summarized as a table in supplement material. 
They are provided for studies of $r$-process as well as other fields such as nuclear engineering. 
Study on level structures and fission barrier of neutron-rich nuclei are highly requested for a further development of the database in future.
\end{abstract}

\pacs{21.60.-n, 21.60.Jz}

\maketitle

\section{Introduction}
\label{Intro}
Origin of chemical elements in the universe is the long-standing problem and one of the hottest topics in astrophysics.
Our knowledge accumulated so far indicates that heavy elements in the nature are generated by dynamical processes of stars. 
In particular, about a half of the elements heavier than iron is considered to be produced by $r$-process~\cite{B2FH,Cameron} (another half is by $s$-process~\cite{B2FH}). 
Recent researches employing observation of gravitational waves produce evidence that a neutron star merger is one of the the possible sites of $r$-process~\cite{Hotokezaka2018,Wanajo2018}. 
However, it is not still concluded where the $r$-process occurs.
\par
The solar $r$-process abundance pattern shows a characteristic mass distribution that has three peaks around $A=80 \sim 90$, $130 \sim 138$ and $195 \sim 208$~\cite{Freiburghaus1999,Arcones2011,Zhi2013}. 
The $r$-process simulation suggests that the origin of these peaks is related to neutron magic numbers of $N=50, 82$ and $126$, that is also related to nuclear mass~\cite{Freiburghaus1999,Langanke2003,Arcones2011}. 
However, the peak positions and abundance ratios cannot be reproduced only by the nuclear mass. 
Neutron capture, $\beta$-decay, and other decay modes sensitively influence the abundance pattern of $r$-process~\cite{Arcones2011,Beun2008,Surman2009,Mumpower2012,Xu2014,Borzov2000,Suzuki2012,Tomislav2016,Mumpower2016b,Nishimura2016,Suzuki2018,Kodama1973,Kodama1975,Kratz1988,Beun2008b,Goriely2015,Eichler2016,Shibagaki2016,Mumpower2018,Giuliani2018}.
\par
During $r$-process, $\beta^-$-decay increases the atomic number of nuclei and brings daughter nuclei in a highly excited state.
Depending on the excitation energy, the daughter nuclei are able to decay through several particle emission channels.
In particular, $\beta$-delayed neutron emission and fission (hereafter we call them BDNE and BDF, respectively) play a subsidiary role in the $r$-process abundance.
BDNE gives two different effects to the $r$-process abundance. 
First is that it leads nuclei in a $r$-process site to detour $\beta^-$-decay path $(A,Z)\rightarrow(A,Z+1)$ by reducing neutron number, for example, $(A,Z)\rightarrow(A-1,Z+1)$ in case of one-neutron emission. 
Another is that BDNE boosts environmental neutron density during the freeze-out phase and stimulates nuclei to capture the neutrons again. 
As a result, the even-odd fluctuation in the final $r$-process abundance pattern is smoothed~\cite{Kodama1973,Kodama1975,Kratz1988,PhysRevC.83.045809}. 
The role of BDF is to reduce abundance of heavy element by breaking daughter nuclei into two or more fragments during $r$-process.
Furthermore, the fission fragments restart to capture environmental neutrons and grow up toward heavy elements again.
This phenomena, what is called fission recycling, affects a wide range of the $r$-process abundance together with neutron-induced fission~\cite{Beun2008b,Goriely2015,Eichler2016,Shibagaki2016,Mumpower2018,Giuliani2018}.
Understanding of the fission in $r$-process would also have a clue to answer a naive question; are super-heavy elements produced by dynamical processes of stars~\cite{Petermann2012}.
\par
A lot of experimental measurements of BDNE are carried out because of its importance in applications to $r$-process as well as nuclear engineering ({\it e.g.} see Refs. ~\cite{Caballero-Folch2016,Rykaczewski2018,Miernik2018,Agramunt2016,Dunlop2019} for recent works).
However, the experimental difficulty rapidly increases as one tries to study very neutron-rich nuclei because of the low statistics.
For this reason, there are still nuclei that the BDNE and the BDF has not been measured yet.  
In particular, for BDF 7 cases are recognized only near the $\beta$-stability line~\cite{ENSDF} (see Ref.~\cite{Ghys2015} which summarizes experimental studies for BDF).
\par
Nuclei that are not investigated by experiment have to be covered by theoretical models.
Empirical systematics is one of the useful tools and is provided in {\it e.g.} Refs.~\cite{Kratz1973,McCutchan2012,Miernik2013,Miernik2014} for $\beta$-delayed neutron branching ratio ($P_n$) and even for BDF in Refs.~\cite{Andreyev2013,Ghys2015}. 
Gross Theory (GT2)~\cite{Yoshida2000,Tachibana2000} that is based on a phenomenological approach is also an effective tool to calculate $P_n$. Although no challenge has been yet performed for delayed fission branching ratio ($P_f$) with GT2, it is technically possible.
%
\par
Another effective approach for theoretical prediction of $P_n$ and $P_f$ is a microscopic model.
A configuration interacting (CI) model, quasiparticle random phase approximation (QRPA), and finite amplitude method (FAM) which is essentially same as the QRPA but solves in a different way, are the typical approaches. Recently, interacting boson model (IBM) based on mean-field approach has been applied to $\beta$-decay calculation as well~\cite{Nomura2020}.
Although CI models are actively applied to the calculation of half-lives for $r$-process nuclei around $N = 50, 82$, and $126$ shell closures, the application is still restricted to limited nuclei due to the increasing computational cost~\cite{Suzuki2012,Yoshida2018,Suzuki2019}.
For this reason, to calculate $\beta$-decay for $r$-process nuclei, the QRPA ~\cite{Staudt1992,Hirsch1992,Moller1997,Borzov2005,Nabi2016,Panov2016} and FAM~\cite{Mustonen2014,Mustonen2016,Ney2020}, are practically applied. 
\par
One of the authors has calculated $P_n$ systematically for neutron-rich nuclei in the framework of proton-neutron relativistic quasiparticle random phase approximation ($pn$-RQRPA)~\cite{Tomislav2016}. 
To predict $\beta$-delayed neutron branching ratios, a simplified approach (referred to as cutoff method hereafter) assuming that nuclei with excitation energies above neutron threshold always emit $\beta$-delayed neutron has been used, as like in Refs.~\cite{Moller1997,Borzov2005,Nabi2016}.
This assumption corresponds with a picture that neutron emission is the only exit channel above neutron thresholds and kinetic energy of $\beta$-delayed neutron is zero.
This prescription clearly omits the nuclear structure, the selection rule of decay chain, competition with other decay channels, and kinematics. 
In fact, it is pointed out in~\cite{Mumpower2016} that a competition between neutron emission and $\gamma$ de-excitation gives a non-negligible variation to the calculation for $P_n$ of nuclei, especially near the neutron drip line.
\par
One of the approaches to treat nuclear decay in a more physical and complete way is to apply a statistical decay model, for example, Hauser-Feshbach statistical model (HFM)~\cite{HFSM}.
The HFM considers nuclear structure effects through level densities and selection rule of decay chain, competition with other decay channels, and kinematics that the cutoff method omitted.
Combination of the QRPA $\beta$ strength function and the statistical decay model is therefore a feasible approach and has been carried out by several groups using FRDM+QRPA~\cite{Kawano2008,Mumpower2018,Moller2019} and non-relativistic QRPA~\cite{Thielemann1983,Staudt1992,Hirsch1992,Panov2005,Minato2016,Panov2016}.
The aim of this paper is directed at estimating $P_n$ and $P_f$ by using this approach, namely by combining $pn$-RQRPA and HFM.
Hereafter, we refer to the present work as $pn$-RQRPA+HFM to distinguish the work of $pn$-RQRPA~\cite{Tomislav2016}.
\par
This paper is organized as follows. 
In Sect.~\ref{sec:theory}, we describe theoretical framework to calculate BDNE and BDF using $pn$-RQRPA+HFM. 
In Sect.~\ref{sec.result}, the results obtained in this work are presented and discussed in comparison with experimental data and preceding works.
Section~\ref{sec:summary} summarizes this work and presents some perspectives.
The complete data table containing the BDNE and BDF branching ratios is available as Supplemental Material~\cite{supplement}.

\section{Theoretical Framework of $pn$-RQRPA+HFM}
\label{sec:theory}
\subsection{$\beta$-delayed neutron and fission branching ratios}
\label{sec:2.A}
Our calculation is composed of two parts. 
First, we prepare $\beta$ strength functions for the Gamow-Teller (GT) and the first-forbidden (FF) transitions by using the $pn$-RQRPA~\cite{Tomislav2016}.
As second step, we carry out the calculation of statistical decay from the compound state by using the HFM calculation with excitation energy and spin-parity given by the $pn$-RQRPA. 
The $P_n$ and $P_f$ are then obtained by multiplying neutron and fission emission probabilities by the $\beta$ decay rates, respectively. 
\par
A fully self-consistent covariant density functional theory (CDFT) framework is adopted in this work.
The ground state of all nuclei is calculated with the relativistic Hartree-Bogoliubov (RHB) model with the D3C* interaction~\cite{Marketin2007}.
The ground state of odd nuclei are computed by employing the same model as for even-even nuclei, but constraining the expectation value of the particle number operator to an odd number of protons and/or neutrons.
On the top of the RHB, excited states are obtained within the $pn$-RQRPA.
More details about the $pn$-RQRPA used in this work are given in Ref.~\cite{Tomislav2016}. 
\par
We assume daughter nuclei reach the compound state, namely the thermally equilibrium state, soon after $\beta^-$-decay. 
At the compound state, daughter nuclei lose initial information that they had before $\beta$-decay, except for the spin-parity and the total energy of the system. 
The number of protons and neutrons of the initial nuclei (precursor) are defined as $Z$ and $N$, respectively. 
Accordingly, the number of protons and neutrons of the daughter nuclei are given by $Z+1$ and $N-1$, respectively. 
Because the $pn$-RQRPA gives excitation energy with respect to the parent nuclei, a correction is required to obtain the energy with respect to the daughter nuclei.
With the BCS approximation, the excitation energies are computed through~\cite{Engel1999, RingandSchuck}
\begin{eqnarray}
E_i^*=&E_{i,QRPA}-E_{corr},\\
E_{corr}&=
\begin{cases}
 E_{p_0}+E_{n_0}& (\mathrm{for\,even-even})\\
 E_{p_0} & (\mathrm{for\,even-odd})\\
 E_{n_0} & (\mathrm{for\,odd-even})\\
 0 & (\mathrm{for\,odd-odd}),
\end{cases}
\label{eq:correct}
\end{eqnarray}
where $E_{i,QRPA}$ is the excitation energy calculated by the $pn$-RQRPA, index $i$ denotes an excited state of daughter nuclei with spin-parity $J^\pi$, and $E_{p_0}$ and $E_{n_0}$ are the lowest quasi-particle energies of proton and neutron, respectively.
\par
We define isotope production ratios of evaporation residues with proton number $Z'$ and neutron number $N'$ as $p_{Z'N'}(i)$, and spectra of emitted particles as $d_\nu(E_\nu,i)$.
Here, $E_\nu$ is the kinetic energy of the outgoing particle and $\nu=\{ n,\gamma,p,\alpha \}$ represents a kind of the emitted particle, where the letters in the brackets represent neutron, $\gamma$-ray, proton, and $\alpha$-particle, respectively. 
We did not consider other particle emissions because they are strongly hindered for neutron-rich nuclei of interest.
The BDF may occur directly after $\beta^-$-decay {\it i.e.} $(\beta^-,f)$ and indirectly after multi-neutron emissions {\it i.e.} $(\beta^-,xnf)$.
The fission probabilities from an excited state $i$ are defined as $p_{xn,f}(i)$.
The functions of $p_{Z'N'}(i)$, $p_{xn,f}(i)$, and $d_\nu(E_\nu,i)$ are computed by the HFM calculation implemented in CCONE code~\cite{CCONE}. 
We do not go into the detail about the HFM because the formalism is given in, {\it e.g.} Refs~\cite{Panov2005,CCONE,Mumpower2016}. 
However, we will explain the nuclear input details used in our calculation in the Sect.~\ref{sec:2.B}.
\par
BDNE branching ratios are calculated by
\begin{equation}
P_{n}=\sum_x P_{xn}=\frac{1}{R}\sum_{i,x} r_i \, p_{Z+1,N-1-x}(i)
\label{eq.branchingPn}
\end{equation}
and BDF branching ratios by
\begin{equation}
P_{f}=\sum_x P_{xn,f}=\frac{1}{R}\sum_{i,x} r_i \, p_{xn,f}(i),
\label{eq.branchingPf}
\end{equation}
where $r_i$ are the partial $\beta^-$-decay rates to excited state $i$ calculated by the $pn$-RQRPA, and $R=\sum_{i}r_i$.
The BDNE spectrum is
\begin{equation}
D_n(E_n)=\mathcal{N}\sum_{i}r_i \, d_n(E_n,i),
\label{eq.dnspectrum}
\end{equation}
where $\mathcal{N}$ is the normalization factor to be determined so as to satisfy
\begin{equation}
\int_0^\infty D_{n}(E_{n}) dE_{n}=1.
\label{normalization}
\end{equation} 
The summation of $i$ of Eqs.~\eqref{eq.branchingPn}, \eqref{eq.branchingPf} and \eqref{eq.dnspectrum} is carried out for $E_i^* \le Q_\beta$.
The $\beta$-decay $Q$ value is calculated from $Q_\beta=M_{n\rm{H}}+\lambda_n-\lambda_p-E_{corr}$~\cite{Tomislav2016} where $M_{n\rm{H}}$, $\lambda_{n}$, and $\lambda_{p}$ are the mass difference between neutron and Hydrogen atom, and the neutron and proton Fermi energies, respectively.

\subsection{Nuclear input details for HFM}
\label{sec:2.B}

\begin{figure}
\includegraphics[width=0.98\linewidth]{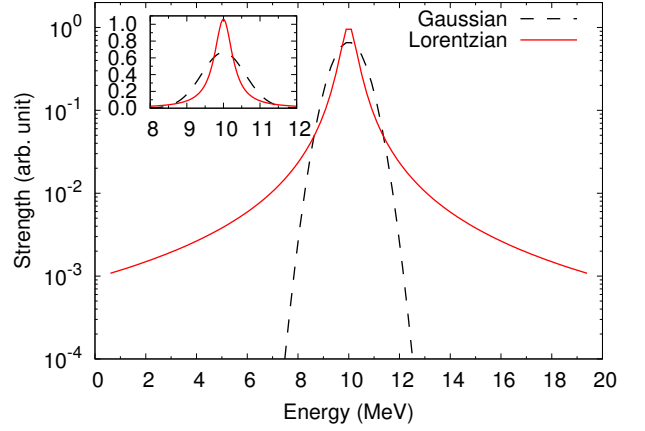}
\caption{Weight functions of Gaussian type (the dotted line) of Eq.~\eqref{eq:Gauss} and Lorentzian (the solid line) type of Eq.~\eqref{eq:smooth_r} with $E^*=10$ MeV and width $\Gamma=0.6$ MeV. The inserted panel is depicted in a linear scale.}
\label{fig:lorentz}
\end{figure}

The $pn$-RQRPA calculation~\cite{Tomislav2016} is solved by diagonalizing RQRPA matrix so that the eigenvalues are given in a form of discrete states. 
Accordingly, $\beta$ strength function also has a discrete shape in terms of excitation energy. 
However, it is considered that actual $\beta$ strength function has a broader distribution because of coupling to higher-order configurations and continuum states, which are not taken into account in the $pn$-RQRPA used in this work. 
In order to account for those effects, we introduce two types of weight function, that is the Gaussian type and the Lorentzian type, to the $\beta$ strength functions.
The weight function of the Gaussian type is given as
\begin{equation}
G_w(E)=g_J\sum_{i \in w} r_i \frac{1}{\sqrt{2\pi\Gamma^2}} \exp\left(-\frac{(E-E_i^*)^2}{2\Gamma^2}\right).
\label{eq:Gauss}
\end{equation}
and that of the Lorentzian type as
\begin{equation}
L_w(E)=g_J\sum_{i \in w} \mathcal{N}_i r_i \frac{1}{\pi} \frac{\Gamma/2}{(E-E_i^*)^2+(\Gamma/2)^2}.
\label{eq:smooth_r}
\end{equation}
The index $w$ is used to distinguish the GT ($\Delta J^\pi=1^+$) and the FF transitions ($\Delta J^\pi=0^-,1^-,2^-$), and the factor $g_J$ is a statistical factor that will be explained later on.
Since the Lorentzian function that is proportional to $1/E^{2}$ has a finite strength even far from the mean, we introduce a cutoff energy $E_{cut}$ to reduce anomalously large strengths at distant energies. 
This cutoff energy is determined by $L_{\omega}(E_{cut})=L_{\omega}(E_i^*)/1000$, and the weight function of Eq.~\eqref{eq:smooth_r} is active within the energy range of $E_i^*\pm E_{cut}$.
The factor $\mathcal{N}_i$ in Eq.~\eqref{eq:smooth_r} is then introduced to renormalize the Lorentzian function to be unity.
In this scheme, a part of $\beta$ strength functions may stray to the negative energies in terms of daughter nuclei.
Because the HFM cannot compute particle evaporations properly if excitation energies are negative, we integrate the $\beta$ strengths at negative energies and set them at $E=0$ MeV, namely, the ground state of the daughter nuclei.
A schematic picture that depicts Eqs.~\eqref{eq:Gauss} and \eqref{eq:smooth_r} with $E^*=10$ MeV and $\Gamma=0.6$ MeV ($E_{cut}=9.5$ MeV) are shown in Fig.~\ref{fig:lorentz}. 
Within $E^*\pm2\Gamma$, the Gaussian function has a broader distribution than the Lorentzian function. 
Beyond this energy range, the Gaussian function rapidly fades out while the Lorentzian function still has a finite strength distribution.
In the next section, we determine the width parameter $\Gamma$ that minimizes the root mean square deviation of $P_{n}$ from the experimental data.
\par
%

%
\par
The $pn$-RQRPA is used in calculations for odd-nuclei in the same way as for even-nuclei, so that the spin-parity of the ground state ($J_{gs}^{\pi_{gs}}$) was not specified. We therefore adopt the experimental data of the spin-parity if it is available. If not, we use the spin-parity of the state with the lowest quasiparticle energy from the RHB calculation, namely $J_{gs}=j_{p}$ with $\pi_{gs}=(-1)^{l_{p}}$ and $J_{gs}=j_{n}$ with $\pi_{gs}=(-1)^{l_{n}}$ for odd-even and even-odd nuclei, respectively, where $j_{p}(j_{n})$ and $l_{p}(l_{n})$ are the quasiparticle total and orbital angular momentum of proton (neutron), respectively. For spin state of odd-odd nuclei, we apply Nordheim method~\cite{Nordheim1950}, which is given by
\begin{equation}
\begin{split}
J_{gs}&=j_p+j_n \,\, \mathrm{if} \,\, j_p=l_p\pm\frac{1}{2} \,\, \mathrm{and} \,\, j_n=l_n\pm\frac{1}{2} \\
J_{gs}&=|j_p-j_n| \,\, \mathrm{if} \,\, j_p=l_p\pm\frac{1}{2} \,\, \mathrm{and} \,\, j_n=l_n\mp\frac{1}{2}
\end{split}
\end{equation}
and the parity is computed via $\pi_{gs}=\pi_{gs}^{(p)}\pi_{gs}^{(n)}=(-1)^{l_p+l_n}$.
The spin of daughter nuclei ($J_f$) resulted from $\beta$-decay of odd-nuclei takes $|J_{gs}-\Delta J| \le J_f \le J_{gs}+\Delta J$ and the parity $\pi_f=\pi_{gs}\pi_\alpha$, where $\Delta J=0, 1, 2$ and $\pi_\alpha=\pm1$ depending on the type of $\beta$ transitions. 
We assumed the equal distribution of $\beta$ strength function for those possible states of daughter nuclei. 
The factor $g_J$ in Eq.~\eqref{eq:smooth_r} is determined from the number of the possible states of the daughter nuclei. 
For example, in case of $1^\pm$ transition from a parent nucleus with $J=3/2(1/2)$ state, $g_{3/2(1/2)}=1/3(1/2)$.
\par
In the HFM calculation, transmission coefficients of nucleons, deuteron, triton, helium-3 are calculated by Koning-Delaroche optical potentials~\cite{KandD2003} and its folding potentials. 
Transmission coefficient of $\alpha$ particle is calculated by the optical potential of Avrigeanu~\cite{Avrigeanu2010}.
For nuclear level densities, the Gilbert-Cameron method \cite{GC} with Mengoni-Nakajima parameter~\cite{MandN} is adopted.
For $\gamma$ strength functions, the enhanced generalized Lorentzian function~\cite{KandU} is used.
Mass data are taken from the global nuclear mass model~\cite{Liu2011}.
%
\par
Transmission coefficients for fission are calculated as follows. 
We assume a double or triple humped parabolic barrier and the barrier penetrability for each barrier is calculated by the formula of Hill-Wheeler equation~\cite{HillWheeler1953}.
The transmission coefficients are obtained by assuming that the fission process occurs through the transition states above the fission barrier. 
All transition states were approximated by the level density formula described above. 
The transmission coefficient of a single barrier for the state having excitation energy $E$ and spin-parity $J^\pi$, $T_i(E,J^\pi)$, is calculated by
\begin{equation}
T_i(E,J^\pi)=\int_0^\infty \frac{\rho_i(\epsilon,J^\pi)}{1+\exp\left(-2\pi\frac{E-V_i-\epsilon}{\hbar\omega_i}\right)}d\epsilon
\end{equation}
where the subscripts $i=A$, $B$, and $C$ indicate the inner, middle, and outer barriers, respectively, $\rho_i(\epsilon,J^\pi)$ the level density at the saddle points, and $V_i$ and $\hbar\omega_i$ represent the height and curvature of the fission barrier, respectively.
The transmission coefficients for two and three barriers is approximated to be $T(E,J^\pi)=T_AT_B(T_A+T_B)$ and $=T_AT_BT_C/(T_AT_B+T_BT_C+T_CT_A)$, respectively.%
\par
Since predicted fission barrier data greatly vary among models, our calculation is carried out using four different fission barriers: HFB-14~\cite{Goriely2007}, FRDM+QRPA~\cite{Moller2015}, Extended Thomas Fermi plus Strutinsky Integral (ETFSI) method~\cite{Mamdouh2001}, and Spherical Basis Method (SBM)~\cite{Koura2014}. 
The latter three give only single barrier information, so that the transmission coefficients are calculated by assuming a single humped parabolic barrier and the barrier curvature~\cite{RIPL3}. 
The curvature parameter we used is $\hbar\omega_A=1.04, 0.80, 0.65$ for even-even, even-odd(odd-even), and odd-odd nuclei, respectively, that are determined to reproduce fission cross sections of uranium isotopes~\cite{CCONE}.
For the case of fission barrier of HFB-14, we adopt the fission barrier data and path given in the database~\cite{Goriely2007}.
\par
\section{Results}
\label{sec.result}

\subsection{$\beta$-delayed neutron emission}
In the last section, the width parameter $\Gamma$ is introduced to make the $\beta$ strength function a broad distribution. 
To determine the most likely $\Gamma$, we estimate the root mean squared (rms) value of $P_{1n}$ defined as
\begin{equation}
\sigma_{\mathrm{rms}}^{xn}=\sqrt{\frac{1}{N}\sum_i^N \left[\log_{10}\left(\frac{P_{xn,i}(c)}{P_{xn,i}(e)}\right)\right]^2},
\label{eq:rms}
\end{equation}
where $N$ is the number of experimental data and $P_{xn,i}(c)$ and $P_{xn,i}(e)$ are the $\beta$-delayed neutron branching ratios of theoretical models and experiment, respectively.
Note that in Ref.~\cite{Mumpower2018} a linear scaled rms is adopted to discuss the predictive power of $P_{1n}$ calculated by the FRDM+QRPA+HFM~\cite{Moller2019}.
However, we adopted logarithmic scaled rms as Eq.~\eqref{eq:rms} because experimental $P_{xn}$ values extend from a small value of order of $10^{-2}$ to $10^2$ as like half-lives.
\begin{figure}
\includegraphics[width=0.98\linewidth]{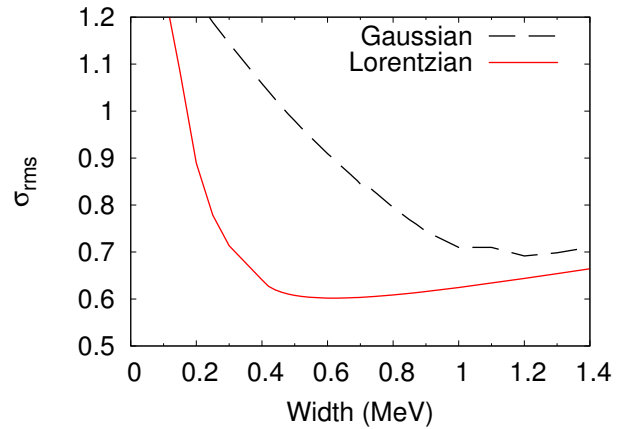}
\caption{Estimated $\sigma_{\mathrm{rms}}^{1n}$ value given in Eq.~\eqref{eq:rms} as a function of the width, $\Gamma$, calculated with the Gaussian function Eq.~\eqref{eq:Gauss} and the Lorentzian function Eq.~\eqref{eq:smooth_r}.}
\label{fig:gamma.pn}
\end{figure} 
\par
Figure~\ref{fig:gamma.pn} shows $\sigma_{\mathrm{rms}}^{1n}$ as a function of width parameter $\Gamma$ of the Gaussian type of Eq.~\eqref{eq:Gauss} (the dashed line) and the Lorentzian type of Eq.~\eqref{eq:smooth_r} (the solid line). 
Experimental data are taken from Ref.~\cite{CRP} and ENSDF~\cite{ENSDF}.
At $\Gamma\sim0$ MeV, $\beta$ strengths feeding to states lower than the BDNE window ($Q_\beta-S_n \le E^*_i \le Q_\beta$) is sizable.
$P_{1n}$ values are mostly underestimated and $\sigma_{\mathrm{rms}}$ is above 1.2 as seen in Fig.~\ref{fig:gamma.pn}.
Increasing width parameter $\Gamma$, $\sigma_{\mathrm{rms}}^{1n}$ monotonically becomes smaller because $\beta$ strengths at energies smaller than neutron separation energy seep into the BDNE energy window and majority of calculated $P_{1n}$ comes close to the experimental values.
Eventually, $\sigma_{\mathrm{rms}}^{1n}$ takes a minimum value at $\Gamma=0.6$ MeV for the Lorentzian type ($\sigma_{\mathrm{rms}}^{1n}=0.602$) and at $\Gamma=1.2$ MeV for the Gaussian type ($\sigma_{\mathrm{rms}}^{1n}=0.694$).
\par
Above the minimal points, $\sigma_{\mathrm{rms}}^{1n}$ turn into a slow increase.
This would be because majority of calculated $P_{1n}$ becomes too large as compared to experimental values.
The Lorentzian type weight function gives the better result than the Gaussian one in terms of $\sigma_{\mathrm{rms}}^{1n}$.
It is difficult to explain this reason because the result of $\sigma_{\mathrm{rms}}^{1n}$ is complicatedly convoluted by many $P_{1n}$ data.
However, we can explain qualitatively that the little leakage of the $\beta$ strength function extending far outside of the mean plays a significant role for a better agreement of $P_{1n}$ with experimental data. 
In fact, the Gaussian type weight function needs a wider width than that of the Lorentzian type to give the minimal $\sigma_{\mathrm{rms}}^{1n}$.
For the subsequent sections in this paper, we thus discuss BDNE and BDF using the weight function of Lorentzian type using $\Gamma=0.6$ MeV,
the width of which is reasonable as comparing to the experimental measurements of low-lying GT resonances for stable tin isotopes investigated by ($^3$He,$t$) reactions~\cite{Pham1995}.
\begin{table*}
\caption{Result of $\sigma_{\mathrm{rms}}^{1n}$ defined by Eq.~\eqref{eq:rms} for different range of $P_{1n}(e)$ (\%).}
\begin{tabular}{c|c|c|c|c|c}
\hline\hline
Model & $P_{1n}(e) <1$ & $1 \le P_{1n}(e) < 10$ & $10 \le P_{1n}(e) < 50$ &$50 \le P_{1n}(e) \le 100$ & All $P_{1n}(e)$ data \\
\hline
$pn$-RQRPA+HFM & $0.952$ & $0.446$ & $0.570$ & $0.317$ & $0.601$\\
$pn$-RQRPA     & $0.857$ & $0.727$ & $0.925$ & $0.460$ & $0.798$\\
GT2            & $0.852$ & $0.656$ & $0.464$ & $0.320$ & $0.595$\\
FRDM+QRPA+HFM  & $1.084$ & $0.442$ & $0.482$ & $0.281$ & $0.512$\\
\hline
Number of nuclei & 50 & 92 & 91 & 34 & 267 \\
\hline
\end{tabular}
\label{tab.sigma}
\end{table*}
\par
Table~\ref{tab.sigma} lists $\sigma_{\mathrm{rms}}^{1n}$ of the $pn$-RQRPA+HFM together with the $pn$-RQRPA~\cite{Tomislav2016}, Gross theory (GT2)~\cite{Yoshida2000, Tachibana2000}, and FRDM+QRPA+HFM~\cite{FRDM+QRPA+HFSM}. 
We also list 4 different $\sigma_{\mathrm{rms}}^{1n}$ classified by range of $P_{1n}$.
The total value of $\sigma_{\mathrm{rms}}^{1n}$ was $0.798$ for the $pn$-RQRPA.
Note that in the $pn$-RQRPA the $\beta$ strength functions are weighted by a Lorentzian function using a width of $65$ keV that is determined so as to reproduce $\beta$-delayed neutron yield of thermal neutron induced fission of $^{235}$U.
The present result of $pn$-RQRPA+HFM greatly improves that of $pn$-RQRPA providing $\sigma_\mathrm{rms}^{1n}=0.601$, which is comparable to that of GT2 ($\sigma_{\mathrm{rms}}^{1n}=0.595$) and FRDM+QRPA+HFM ($0.512$).
Especially, we obtained a remarkable improvement in $\sigma_{\mathrm{rms}}^{1n}$ throughout from $1 \le P_{1n} \le 100$.
Although $\sigma_{\mathrm{rms}}^{1n}$ is deteriorated in $P_{1n} < 1$, the result of $pn$-RQRPA+HFM is still slightly better than FRDM+QRPA+HFM.
\par
\begin{figure}
\includegraphics[width=0.95\linewidth]{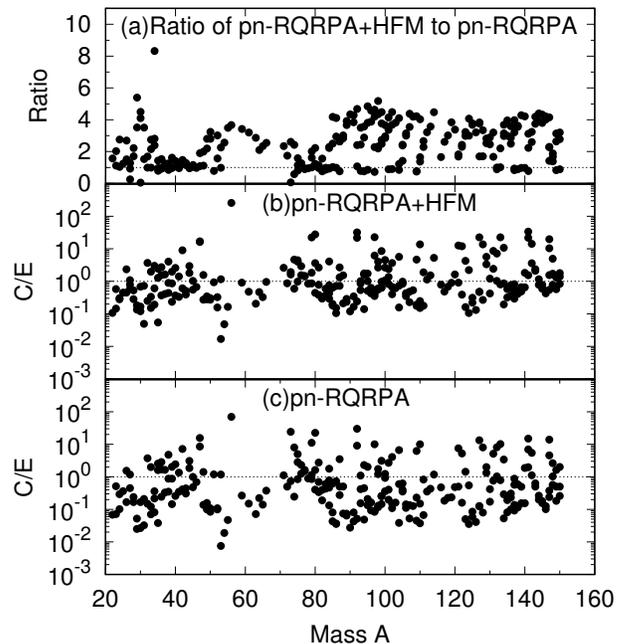}
\caption{
(a) Ratio of $P_{1n}$ for $pn$-RQRPA+HFM to that for $pn$-RQRPA. The panels (b) and (c) represent ratio of calculated to experimental data of $P_{1n}$ ($\mathrm{C/E}$) for $pn$-RQRPA+HFM (this work) and $pn$-RQRPA~\cite{Tomislav2016}, respectively.}
\label{fig:CoE}
\end{figure}

Figure~\ref{fig:CoE}(a) shows the ratio of $P_{1n}$ for $pn$-RQRPA+HFM to that for $pn$-RQRPA as a function of mass number $A$.
We can see that most of data point is above unity, i.e. most of $P_{1n}$ is increased by the framework discussed in Sect.~\ref{sec:theory}.
Figure~\ref{fig:CoE}(b) and (c) shows the ratio of calculated $P_{1n}$ to experimental one ($\mathrm{C/E}$) for $pn$-RQRPA+HFM and $pn$-RQRPA, respectively.
We can observe that underestimations (i.e. $\mathrm{C/E} < 1$) for the $pn$-RQRPA are improved for the $pn$-RQRPA+HFM.
This is mainly due to adopting a wide width of the weight function and inclusion of $n$--$\gamma$ competition which enhances $P_{1n}$.
Although there exists $\beta$ strength function that escapes from one-neutron emission energy window to more multiple ones, this outflow is generally smaller than the gain to one-neutron emission energy window because $\beta$-decay rates become lower as going to higher excitation energies.
We confirmed that $73\%$ of $P_{1n}$ corresponding to $193$ nuclei are improved for the $pn$-RQRPA+HFM.
\par
\begin{figure*}
\includegraphics[width=0.9\linewidth]{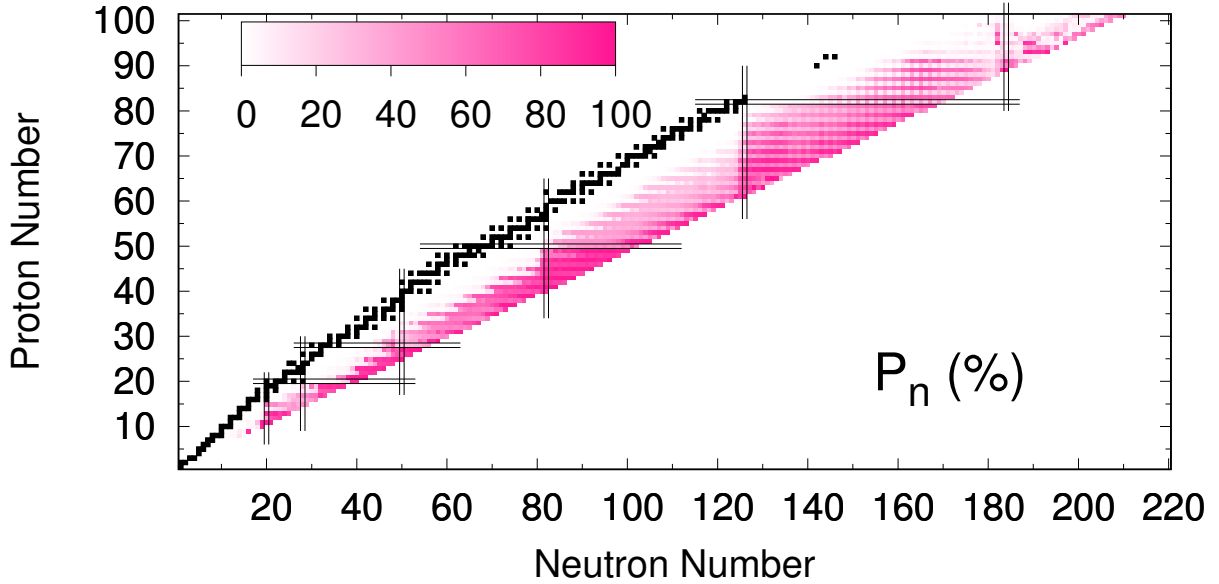}
\caption{Total $\beta$-delayed neutron branching ratios ($P_{n}$) calculated by the $pn$-RQRPA+HFM. The black filled squares stand for stable or long-lived nuclei.}
\label{fig:pnmap}
\end{figure*}

Figure~\ref{fig:pnmap} plots $P_n$ calculated by the $pn$-RQRPA+HFM in the $N$--$Z$ plane. 
We can observe that $P_n$ has a large value in the lower-right sectors seeing from the double magic numbers.
This is simply explained by small neutron threshold energies and large $Q_\beta$ in those regions.
We can also observe the odd-even dependence, especially in $Z$-direction.
Generally, $P_{n}$ of odd-$Z$ nuclei are larger than neighboring even-$Z$ nuclei.
This is because $Q_\beta$ of odd-$Z$ nuclei are larger than those of even-$Z$ nuclei in general, while $S_{xn}$ are less sensitive to $Z$ numbers than $Q_\beta$.
As a consequence, BDNE energy window has an odd-even structure.
As an example, one-neutron emission energy window of $^{220}$Ir ($Z=77$) is about $8.04$ MeV, while that of the neighboring nuclei $^{219}$Os ($Z=76$) and $^{221}$Pt ($Z=78$) are about $7.40$ MeV and $5.65$ MeV, respectively.
Going toward heavy mass region around $Z>90$ and $N>182$, $P_n$ displays a patchy pattern.
This is because BDF plays a role in this region and competes with BDNE (see Sect.~\ref{sec:bdf}).
\begin{figure*}
\centering
\includegraphics[width=0.48\linewidth]{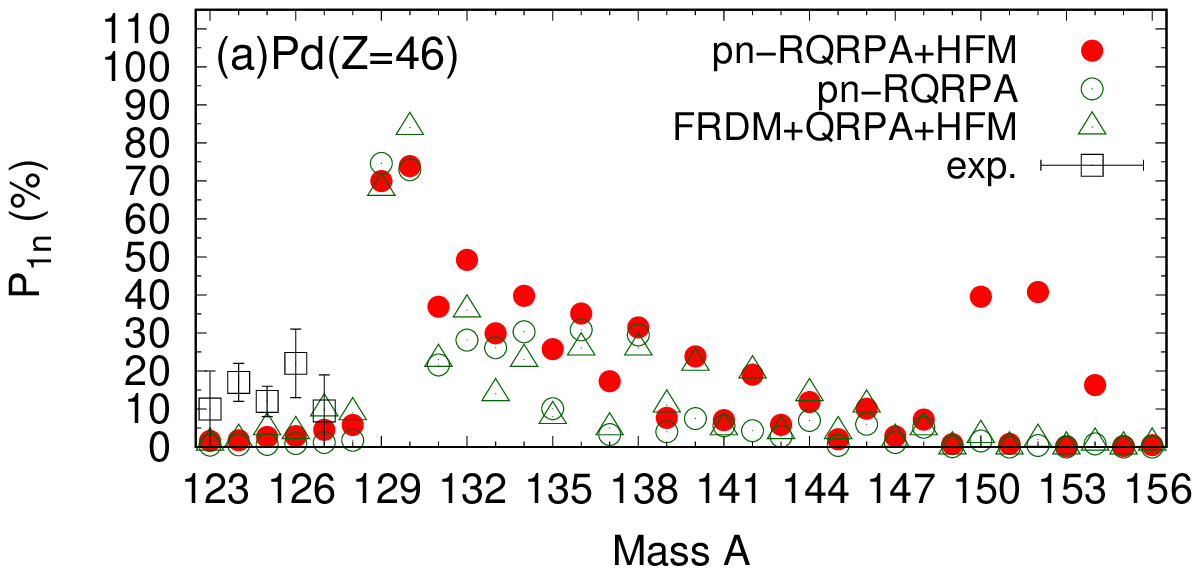}
\includegraphics[width=0.48\linewidth]{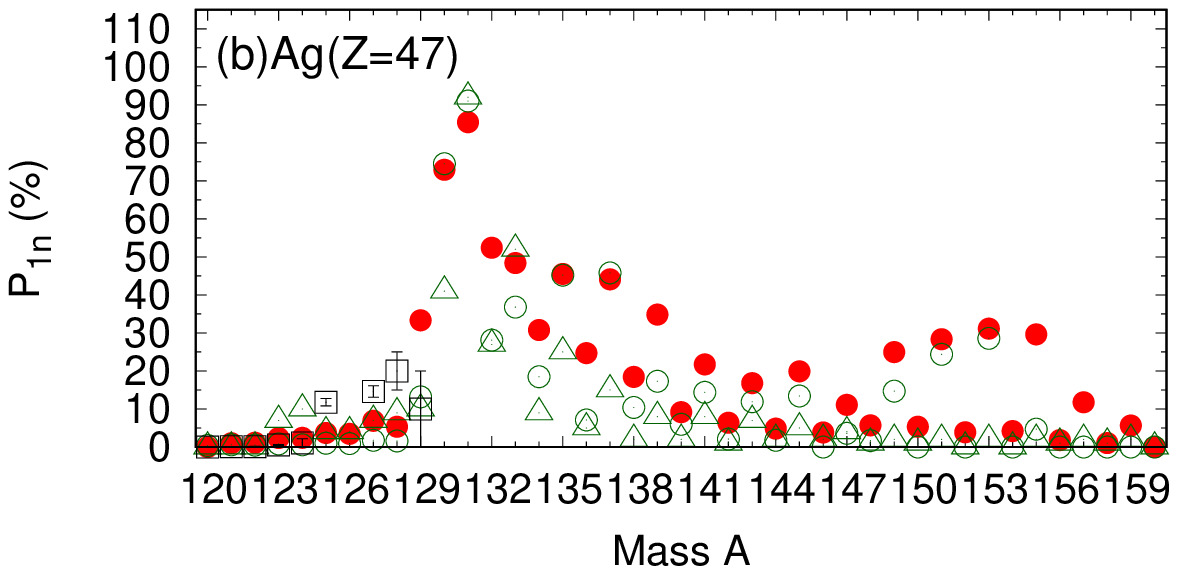}\\
\includegraphics[width=0.48\linewidth]{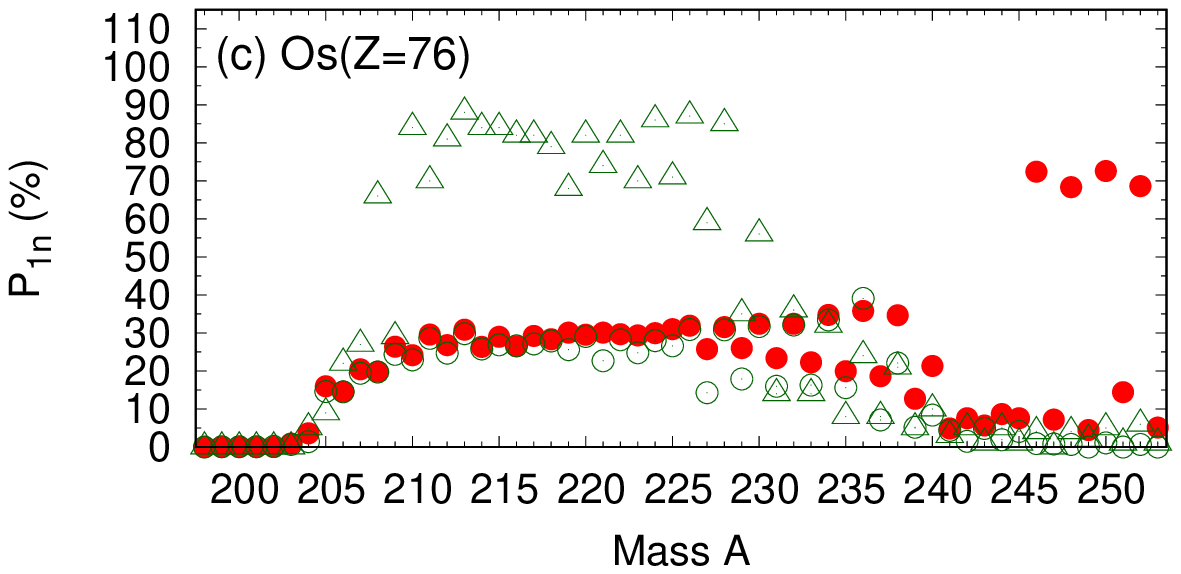}
\includegraphics[width=0.48\linewidth]{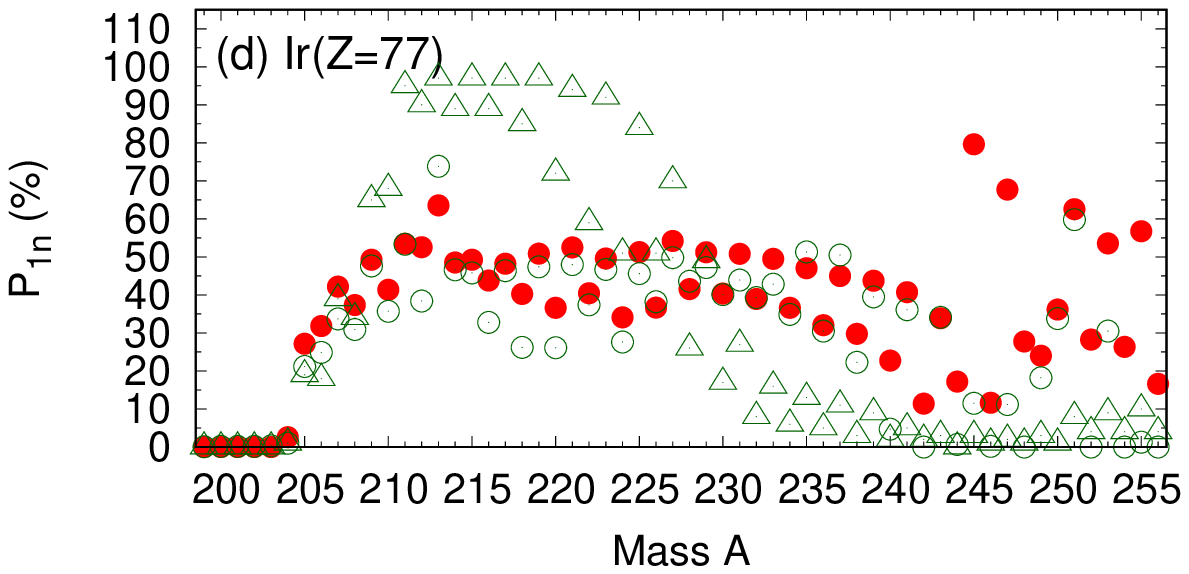}
\caption{$\beta$-delayed neutron branching ratio $P_{1n}$ of Pd, Ag, Os, and Ir isotopes as a function of $A$. The experimental data are taken from Ref.~\cite{CRP}.}
\label{fig:pn.rp}
\end{figure*}
\begin{figure}
\centering
\includegraphics[width=0.95\linewidth]{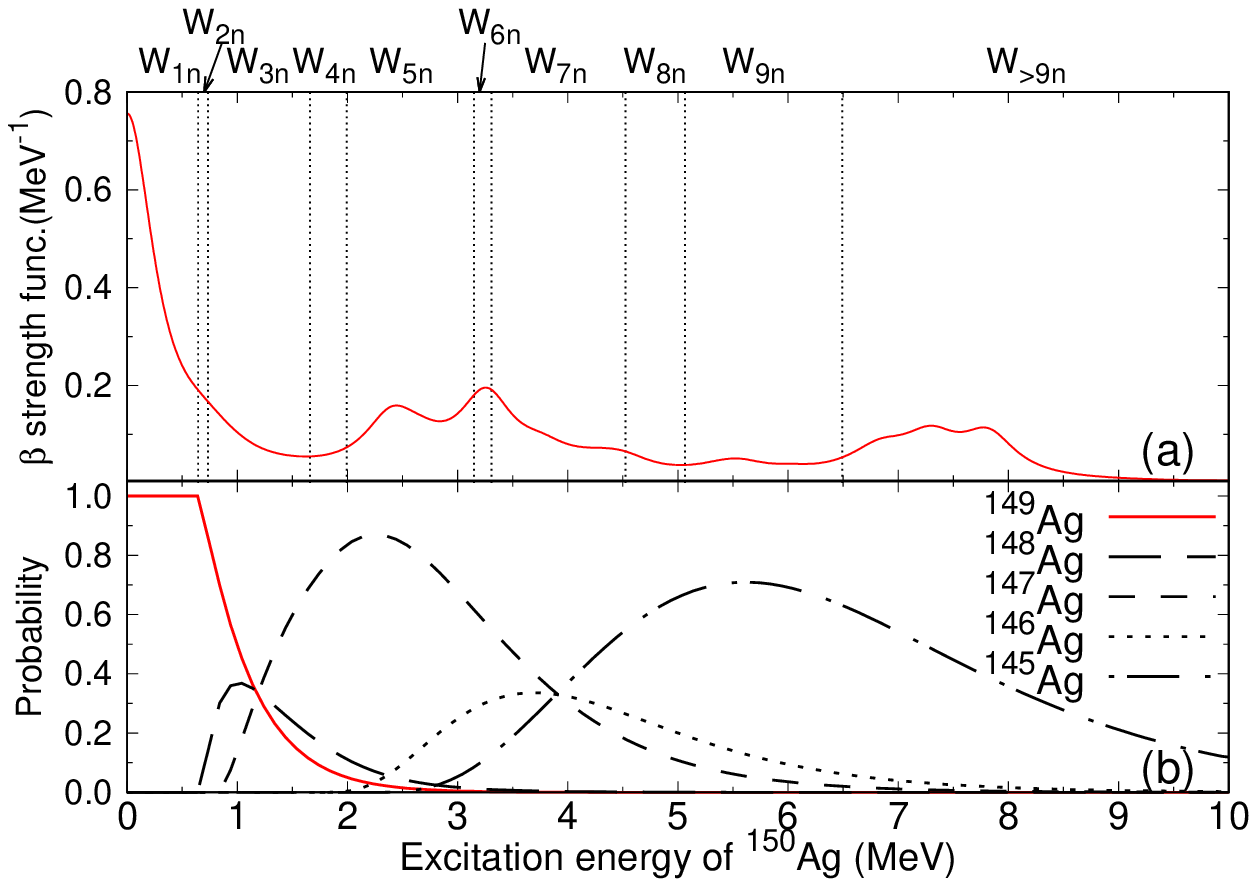}
\caption{
(a) $\beta$ strength function of $^{150}$Pd. BDNE windows denoted by $W_{xn}$ are also indicated. 
(b) Isotope production ratios of $^{145-149}$Ag from $1^+$ excited states of $^{150}$Ag (the daughter nucleus of $^{150}$Pd) as a function of the excitation energy. 
The calculation is performed by the HFM.}
\label{fig:BS}
\end{figure}
\par
Figure~\ref{fig:pn.rp} shows $P_n$ of Pd ($Z=46$), Ag ($Z=47$) isotopes that are considered as important $\beta$-delayed neutron precursors in $r$-process~\cite{Mumpower2016b}, and Os ($Z=76$) and Ir ($Z=77$) isotopes.
We plot the results of $pn$-RQRPA~\cite{Tomislav2016} and FRDM+QRPA+HFM~\cite{Moller2019} for comparison. 
Because the same $\beta$ strength function except the width factor is used for the $pn$-RQRPA+HFM and the $pn$-RQRPA, they give a similar result until a certain mass number $A$.
However, the $pn$-RQRPA+HFM and the $pn$-RQRPA begin to show a difference for neutron-rich side.
For example, $P_{1n}$ of $^{150}$Pd for the $pn$-RQRPA+HFM is about 40\%, while that for the $pn$-RQRPA is $1.6\%$. 
Because the daughter nucleus $^{150}$Ag is an unbound nucleus according to the global nuclear mass model~\cite{Liu2011}, all the $\beta$-decay must contribute to BDNE.
However, a large amount of the $\beta$ strength function of $^{150}$Pd distribute in negative energy in the framework of $pn$-RQRPA~\cite{Tomislav2016}, which means that some $\beta$ strength functions exist at lower energies than the ground state of $^{150}$Ag.
As a result, $P_{0n}$ of $^{150}$Pd unsuitably becomes $35\%$ for the $pn$-RQRPA~\cite{Tomislav2016}, which contradicts the fact that $^{150}$Ag is unbound.
For the present work ($pn$-RQRPA+HFM), the problem of negative energy resonances is avoided by resetting it to zero energy as explained in Sect.~\ref{sec:2.B}. 
As a result, the calculated $P_{0n}$ of $^{150}$Pd is correctly zero and $P_{1n}$ increases by almost the same amount of $P_{0n}$ for the $pn$-RQRPA.
\par
The result of $pn$-RQRPA+HFM shows a similar $P_{1n}$ to FRDM+QRPA+HFM for nuclei with small mass numbers, as shown in Fig.~\ref{fig:pn.rp}. 
As going toward heavier mass, two calculations begin to show a difference.
The noticeable contrasts are found in nuclei close to the neutron drip-line and $210 \le A \le 230$ for Os and Ir isotopes.
A major factor affecting $P_{1n}$ is the $\beta$ strength function and one neutron separation energies, $S_{1n}$, which are calculated from the mass data.
The $pn$-RQRPA+HFM and FRDM+QRPA+HFM use theoretically predicted mass data that is the global nuclear mass model~\cite{Liu2011} and FRDM1995~\cite{Moller1995}, respectively.
For most of nuclei close to the drip line, $S_{1n}$ of the global nuclear mass model is lower than those of FRDM, for example, $S_{1n}$ of $^{240}$Ir is $269$ keV for the global mass model while $690$ keV for FRDM1995.
To check the sensitivity of $P_{1n}$ to mass data, we carry out the same calculation replacing the mass data of the $pn$-RQRPA+HFM with FRDM1995, and find that the result globally becomes close to that of FRDM+QRPA+HFM for nuclei near the drip-line.
However, the deviations found in $210 \le A \le 230$ for Os and Ir isotopes do not change significantly.
We thus consider that the difference found in $210 \le A \le 230$ for Os and Ir isotopes is mainly attributed to the $\beta$ strength function and that found near the neutron drip-line is due to the mass data.
We observe an odd-even staggering of $P_{1n}$ in nuclei near neutron-drip lines for the $pn$-RQRPA+HFM.
This structure appears particularly when daughter nuclei are unstable against neutron emissions.
%
\par
Figure~\ref{fig:BS}(a) shows the $\beta$ strength function of $^{150}$Pd with the Lorentzian function with $\Gamma=600$ keV, in which the BDNE energy windows of $x$ neutron emission (defined as $W_{xn}$) are also drawn by the dotted lines.
In case of the cutoff method, $P_{xn}$ can be directly calculated from the amount of $\beta$-feeding to $W_{xn}$ in Fig.~\ref{fig:BS}.
Figure~\ref{fig:BS}(b) shows the isotope production ratios ($p_{Z+1,N-1-x}(i)$) of Ag isotopes ($149\le A \le 145$) from the $1^+$ excited states of $^{150}$Ag.
As we can see, $p(i)$ distribute into wide range excitation energies and are not stuffed inside a specific BDNE energy window.
Note that $p(i)$ of $^{145}$Ag distributes from $W_{7n}$ to $W_{>9n}$ according to the HFM calculation, while $W_{5n}$ ranges from $2.0$ to $3.1$ MeV.
This mismatch occurs because $\beta$-delayed neutrons take out the excitation energy of daughter nuclei.
On the other hand, the cutoff method assumes zero kinetic energy of emitted neutrons.
Calculated average kinetic energy of $\beta$-delayed neutron for $^{150}$Pd precursor is about $640$ keV.
If 5 neutrons are emitted, about $3.2$ MeV are taken away by the neutrons from daughter nuclei, and this energy is approximately consistent to the difference between $W_{5n}$ and high regions of the survival probability of $^{145}$Ag.
The result of $\beta$-delayed neutron branching ratios for the $pn$-RQRPA+HFM and the cutoff method are listed in Table~\ref{tab:pxn}.
For $x\ge7$, $P_{xn}$ of the cutoff method are larger than the $pn$-RQRPA+HFM results.
The high fractions of $P_{xn}$ of the cutoff method are redistributed to $P_{xn}$ for $x<7$ of the $pn$-RQRPA+HFM.
\begin{table}
\centering
\begin{tabular}{c|c|c}
\hline\hline
(\%) & $pn$-RQRPA+HFM & Cutoff method \\
\hline
$P_{1n}$ & $39.5$ & $27.6$ \\
$P_{2n}$ & $2.0$ & $1.6$ \\
$P_{3n}$ & $20.5$ & $8.1$ \\
$P_{4n}$ & $8.3$ & $2.0$ \\
$P_{5n}$ & $17.9$ & $15.4$ \\ 
$P_{6n}$ & $3.6$ & $3.1$ \\
$P_{7n}$ & $7.1$ & $12.4$ \\
$P_{8n}$ & $0.6$ & $2.5$ \\
$P_{9n}$ & $0.4$ & $6.2$ \\
$P_{>9n}$ & $0.0$ & $21.1$ \\
\hline
\end{tabular}
\caption{$\beta$-delayed neutron branching ratios of $^{150}$Pd.}
\label{tab:pxn}
\end{table}
\par
We also computed delayed neutron yield of thermal neutron induced fission of $^{235}$U. 
The result is shown in Table~\ref{tab:dny}. 
The fission fragment yields of thermal neutron induced fission of $^{235}$U are taken from JENDL Fission Product Yield 2011 (JENDL/FPY2011)~\cite{FPY2011,Katakura2015}. 
The $pn$-RQRPA gives the closest $\beta$-delayed neutron yield to the experimental data among four models because the $P_n$ values are tuned to reproduce it.
The result of $pn$-RQRPA+HFM also reproduces the experimental data in the same order. 
Note that the calculated delayed neutron yields are an aggregated value summed over fission fragment yields.
Looking into important precursor nuclei contributing the delayed neutron yield, the result of $pn$-RQRPA is not necessarily correct.
For example, a precursor contributing the delayed neutron yield the most is $^{137}$I ($P_{1n}(e)=7.66\%$~\cite{ENSDF}) according to the recent evaluation~\cite{CRP}, while it is $^{91}$Rb for the $pn$-RQRPA calculation and $^{137}$I enters the eighth place with $P_{1n}=0.7\%$.
On the other hand, the $pn$-RQRPA+HFM calculation shows $^{137}$I to be the most important precursor, however, the $P_n (\sim 2.25\%)$ is still underestimated and thus the calculated delayed neutron yield is smaller than the experimental value.
\begin{table}
\centering
\caption{$\beta$-delayed neutron yield of thermal neutron induced fission of $^{235}$U. Fission fragment yields are taken from JENDL/FPY-2011~\cite{FPY2011,Katakura2015}.}
\begin{tabular}{c|c}
\hline\hline
Model & $\beta$-delayed neutron yield \\
\hline
$pn$-RQRPA      & $1.43\times10^{-2}$ \\
$pn$-RQRPA+HFM & $1.00\times10^{-2}$ \\
GT2             & $0.81\times10^{-2}$ \\
FRDM+QRPA+HFM  & $0.81\times10^{-2}$ \\
\hline
 exp.~\cite{Keepin1957} & $(1.58\pm0.05) \times10^{-2}$ \\
\hline
\end{tabular}
\label{tab:dny}
\end{table}
\par
\begin{figure*}
\centering
\includegraphics[width=0.46\linewidth]{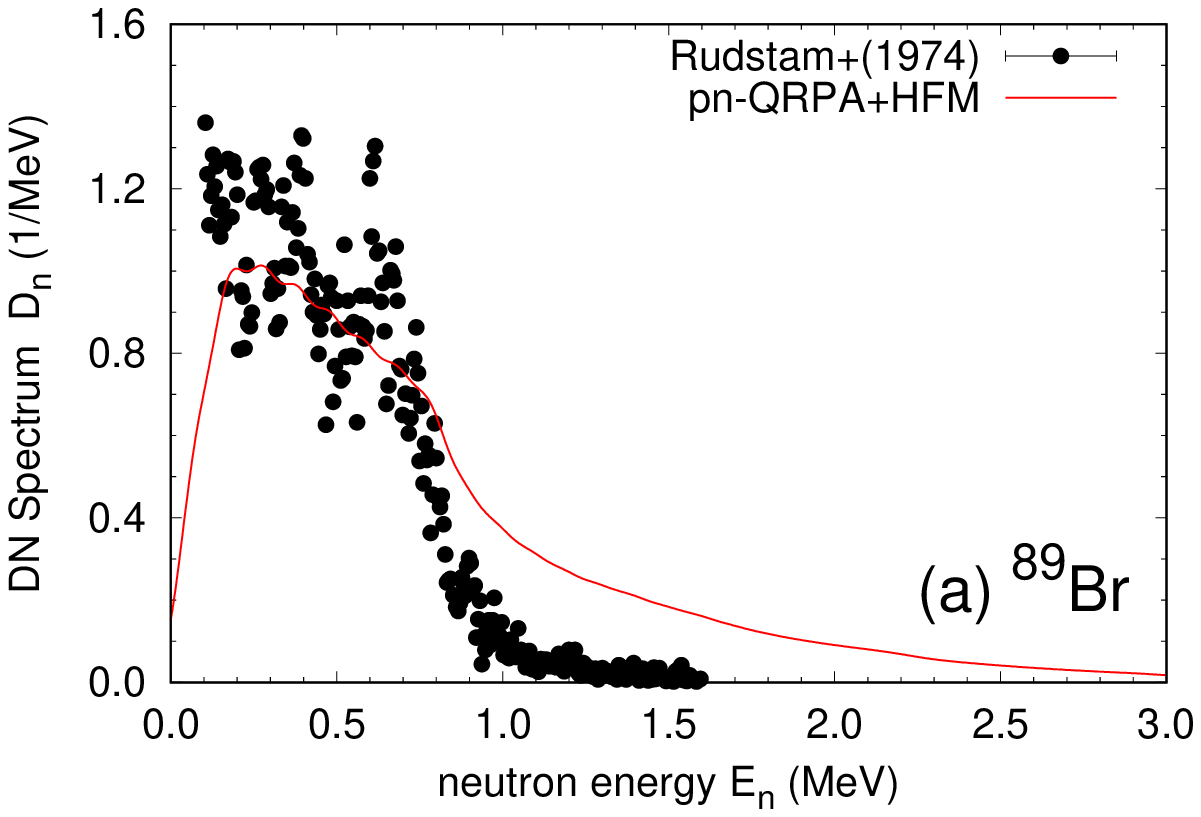}
\includegraphics[width=0.46\linewidth]{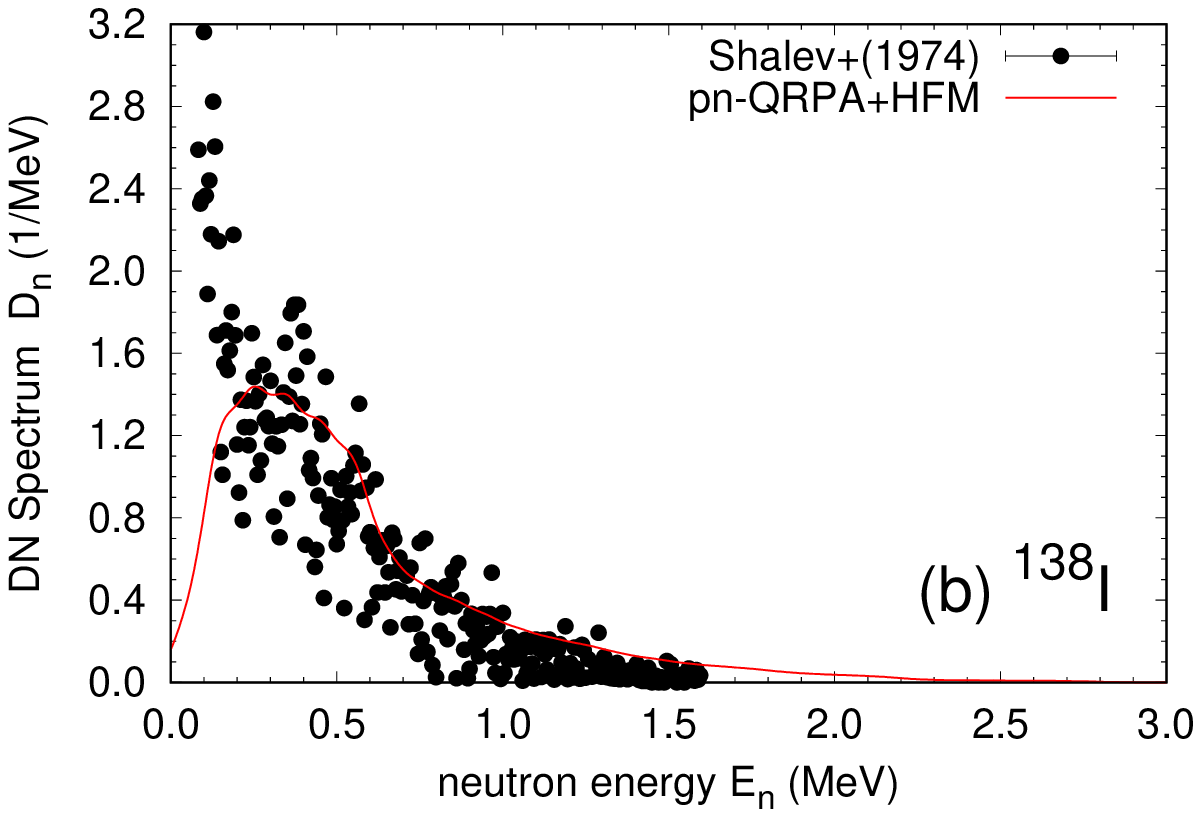}
\caption{BDNE spectrum of (a) $^{89}$Br and (b) $^{138}$I.
Calculated result is shown by the solid line (red). Experimental data are taken from Rudstam et al.~\cite{Rudstam1974} and Shalev et al.~\cite{Shalev1974}.}
\label{fig:nspec}
\end{figure*}

The advantageous to couple the Hauser-Feshbach statistical model is that one can calculate spectra of emitted particles.
As an example, Fig.~\ref{fig:nspec} shows the BDNE spectrum of $^{89}$Br and $^{138}$I, which are typical $\beta$-delayed neutron precursors.
We also plot the experimental data taken from~\cite{Rudstam1974, Shalev1974}.
Although fine structures observed in the experimental data are not reproduced well by the $pn$-RQRPA+HFM, the computed results emulates the experimental $\beta$-delayed neutron spectra reasonably.
\subsection{$\beta$-delayed fission}
\label{sec:bdf}
\begin{figure}
\includegraphics[width=1.0\linewidth]{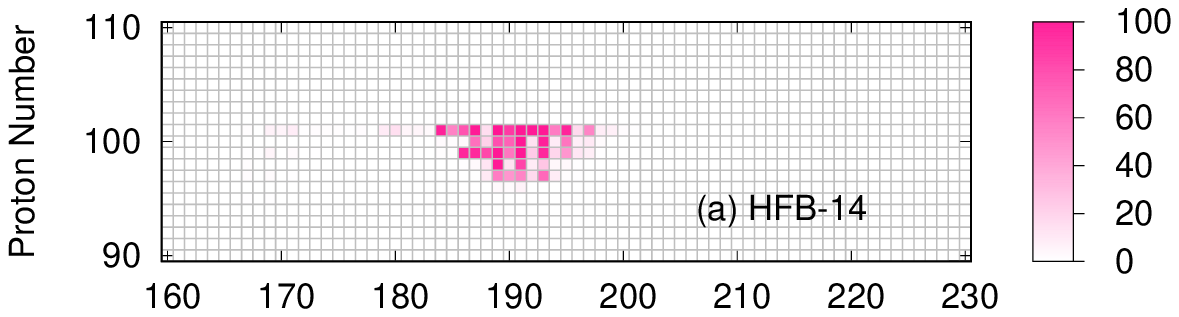}
\includegraphics[width=1.0\linewidth]{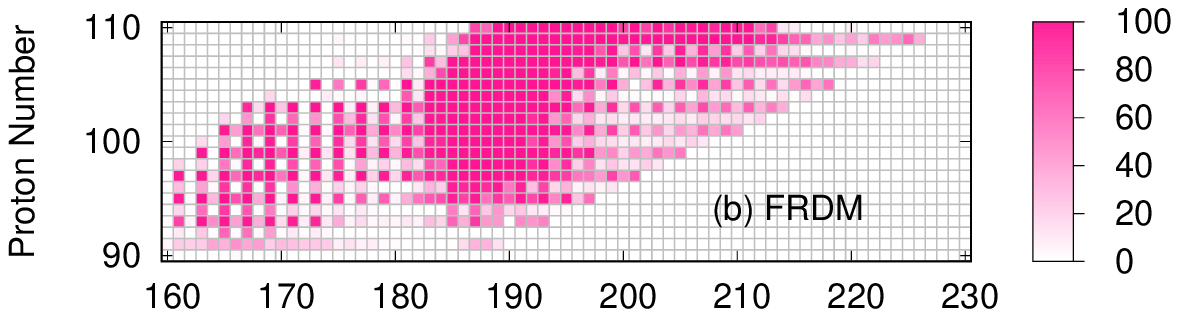}
\includegraphics[width=1.0\linewidth]{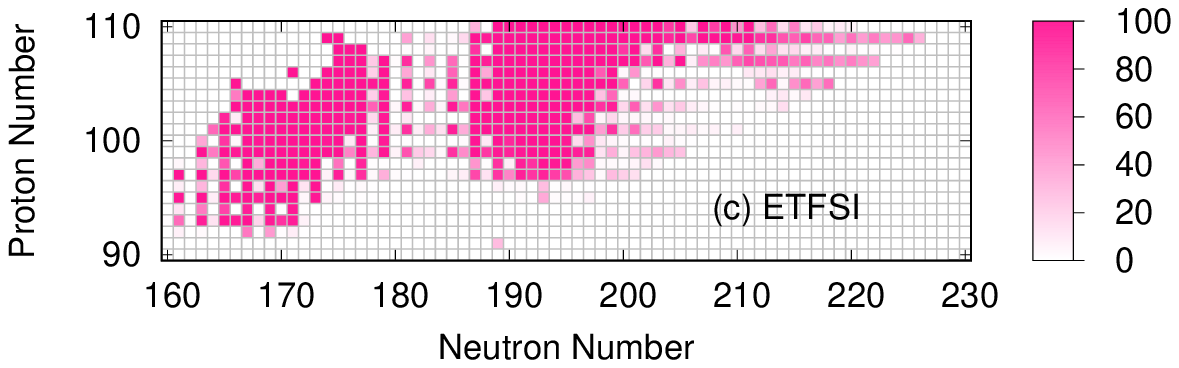}
\includegraphics[width=1.0\linewidth]{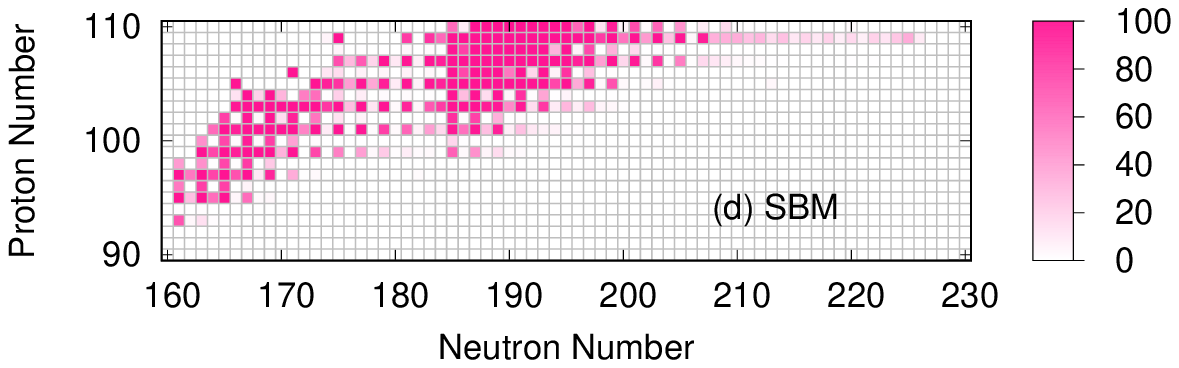}
\caption{BDF branching ratios $P_f$ (\%) calculated by the fission barrier data of (a)HFB-14~\cite{Goriely2007}, (b)FRDM~\cite{Moller2015}, (c)ETFSI~\cite{Mamdouh2001}, and (d)SBM~\cite{Koura2014}. Because HFB-14 provides fission barrier data to $Z=102$, $P_f$ values for $Z\ge103$ are given as blank.}
\label{fig:pfmap}
\end{figure}
Figure~\ref{fig:pfmap} shows BDF branching ratios in the $N$-$Z$ plane using different fission barrier data of (a)HFB-14~\cite{Goriely2007}, (b)FRDM~\cite{Moller2015}, (c)ETFSI~\cite{Mamdouh2001}, and (d)SBM~\cite{Koura2014}. 
Note that the HFB-14 provides fission barrier data only up to $Z=102$, so that $P_f$ over $Z=102$ are given as blank.
Around $101\le Z \le 110$ and $184\le N \le 200$, BDF plays a meaningful role commonly for fission barrier data of FRDM, ETFSI, and SBM.
Although calculated result is limited for the HFB-14 barrier, $P_f$ become meaningful around $Z=100$ and $N=190$ as well.
BDF is also significant from $Z=91, N=160$ to $Z=100, N=185$ for FRDM, ETFSI, and SBM, while that for HFB-14 is negligibly small because its fission barriers are higher than the others.
There exists odd-even dependence in the region of $N=160$ to $185$, in particular, for FRDM and SBM.
This odd-even dependence is also found in Ref.~\cite{Mumpower2018} that calculates $P_f$ with the FRDM+QRPA and the FRDM barrier data~\cite{Moller2015}.
An effect of the neutron magic number of $N=184$ is seen for ETFSI and a valley of $P_f$ is formed, while that for HFB-14, FRDM, SBM are not clearly seen.
\par
BDF branching ratios are correlated to $\beta$-delayed neutron branching ratio. Therefore, we prepared four kinds of data tables depending on fission barrier data, which are available in the supplemental material. 
Undoubtedly $P_n$ are identical among the data tables unless BDF comes into effect.
\subsection{Delayed $\alpha$ emission}
\begin{figure}
\centering
\includegraphics[width=1.0\linewidth]{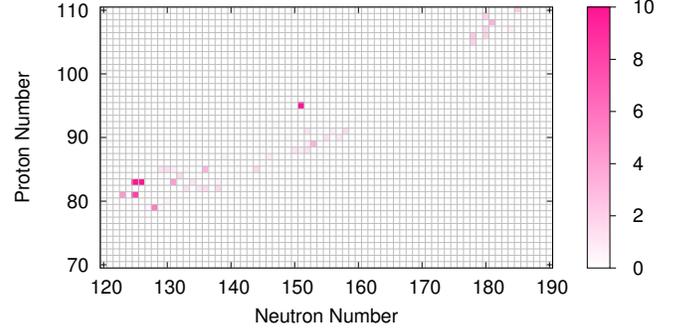}
\caption{Delayed alpha branching ratios $P_\alpha$ (\%) in case of FRDM fission barrier data.}
\label{fig:pamap}
\end{figure}

In this work, we allow daughter nuclei also to decay by $\alpha$-particle emission, and study the $\beta$-delayed $\alpha$ emission branching ratio, $P_\alpha$.
The calculated $P_\alpha$ are shown in Fig.~\ref{fig:pamap} in case of FRDM fission barrier data. 
Note that the maximum scale is set to be $P_\alpha=10$\% for illustration.
Since neutron emission overcomes $\alpha$-particle emission in neutron-rich region, a non-negligible $P_\alpha$ is observed only in a band of near the $\beta$ stability line.
To our knowledge, experimental data on $\beta$-delayed $\alpha$-particle emission has been reported only for $^{214}$Bi ($P_\alpha=0.003$\%)~\cite{NDS210} in the range of Fig~\ref{fig:pamap}.
The calculated result of $P_\alpha$ for $^{214}$Bi is $0.026$\%. 
Although our model overestimates the experimental data, it shows that the $\beta$-delayed $\alpha$-particle emission branching ratio is very small.
\par
Exceptional cases exceeding $P_\alpha=10$\% are $^{210,211}$Bi($Z=83$) and $^{248}$Am($Z=95$), that have $P_\alpha=92.5, 48.9$ and $15.9$\%, respectively.
The daughter nuclei, $^{210,211}$Po and $^{248}$Cm, have a relatively high $\alpha$ decay rates from the ground states~\cite{NDS210}.
It is not thus surprising that they do $\alpha$-particle emission following the $\beta$-decay, competing with $\gamma$ and other de-excitations.
We checked delayed proton emission branching ratios $P_p$ as well.
However, no significant $P_p$ is observed in this work. 
Note that our calculation is carried out only for nuclei in the $\beta$-stability line to a neutron-rich region.
Our future plan is to study a neutron-deficient region where it is expected that $P_\alpha$ as well as $P_p$ become more important than $P_n$.

\section{Summary}
\label{sec:summary}
In this work, we calculated the BDNE and BDF by combining the discrete $\beta$ strength function provided by the $pn$-RQRPA~\cite{Tomislav2016} and Hauser-Feshbach statistical model~\cite{HFSM}. 
We obtained the root mean square of $P_{1n}$ in a good agreement with experimental data, which was comparable to the results of proceeding works, and could improve $P_{1n}$ in a wide range of nuclei as compared to Ref.~\cite{Tomislav2016}.
We also discussed the different role of Gaussian type and the Lorentzian type weight function.
\par
For nuclei near the neutron drip line, we concluded that $P_{1n}$ of the $pn$-RQRPA were largely different from that of the $pn$-RQRPA+HFM. 
We explained that these deviations are attributed to the problem of negative resonance energies resulted from the correction in Eq.~\eqref{eq:correct} and the energy withdrawal from daughter nuclei by $\beta$-delayed neutrons. 
Calculated $\beta$-delayed neutron spectra of $^{89}$Br and $^{138}$I were compared with the experimental data and it turned out that the $pn$-RQRPA+HFM was able to reasonably reproduce the experimental data.
We also calculated the $\beta$-delayed neutron yield of thermal neutron induced fission of $^{235}$U.
The computed result was in an agreement with the experimental data in the same order.
\par
BDF branching ratios are calculated with four different fission barrier data.
We observed a strong dependence of $P_f$ on the fission barrier data used.
By comparing the result of different fission barrier data, BDF may particularly become important around $101\le Z \le 110$ and $184\le N \le 200$.
Since fission barrier data have a large uncertainty, a further study is highly required to investigate $P_f$ as well.
\par
We have to mention that level structures of neutron-rich nuclei are calculated by the phenomenological method described in Sect.~\ref{sec:2.A} because they are not known.
A more accurate data of BDNE and BDF is expected if the level structures would be investigated well.
Further study on this respect is required.
\par
A table of BDNE, BDF, and $\beta$-delayed $\alpha$-particle emission branching ratios calculated in this work is available in the supplement material~\cite{supplement}.

\begin{acknowledgments}
N.P. acknowledges support by the QuantiXLie Centre of Excellence, a project co financed by the Croatian Government and European Union through the European Regional Development Fund, the Competitiveness and Cohesion Operational Programme (KK.01.1.1.01).
\end{acknowledgments}
\nocite{*}
\bibliography{biblio}

\end{document}